\documentclass[]{piparticle-final}
\usepackage{graphicx}
\usepackage{amsmath}

\usepackage{cite} 
\usepackage{epstopdf}

\begin{document}

\volume{6}               
\articlenumber{060001}   
\journalyear{2014}       
\editor{P. Weck}   
\reviewers{J. P. Marques, Departamento de F\'isica, Centro de F\'isica At\'omica, \\ \mbox{}\hspace{41mm} Fac. de Ci\^encias, Universidade de Lisboa, Portugal.}  
\received{19 December 2013}     
\accepted{20 January 2014}   
\runningauthor{E. V. Bonzi \itshape{et al.}}  
\doi{060001}         

\title{Experimental determination of $L$ X-ray fluorescence cross sections for elements with 45 $\leq$ Z $\leq$ 50 at 10 keV}

\author{E. V. Bonzi,\cite{inst1,inst2}\thanks{E-mail: bonzie@famaf.unc.edu.ar}\hspace{2mm}  
        G. B. Grad,\cite{inst1}\thanks{E-mail: grad@famaf.unc.edu.ar}\hspace{2mm}
        R. A. Barrea\cite{inst3}\thanks{E-mail: rbarrea@depaul.edu}}

\pipabstract{
Synchrotron radiation at 10 keV was used to experimentally determine the $Ll$, $L\alpha$, $L\beta_I$, $L\beta_{II}$, $L\gamma_I$ and $L\gamma_{II}$ fluorescence cross sections for elements with  45 $\leq$ Z $\leq$ 50, as part of an ongoing investigation at low energies.
The measured data were compared with calculated values obtained using coefficients from Scofield, Krause and Puri {\it{et al.}} 
}

\maketitle
\blfootnote{
\begin{theaffiliation}{99}
   \institution{inst1} Facultad de Matem\'atica, Astronom\'{\i}a y F\'isica, Universidad Nacional de C\'ordoba. Ciudad Universitaria. 5000 C\'ordoba, Argentina.
   \institution{inst2} Instituto de F\'{\i}sica Enrique Gaviola (CONICET), 5000 C\'ordoba, Argentina.
   \institution{inst3} Physics Department, DePaul University, Chicago, IL 60614, USA.
\end{theaffiliation}
}

\section{Introduction}
This work is part of a systematic investigation on elements with 45 $\leq$ Z $\leq$ 50, which has been carried out at different 
energies \cite{Bonzi2005, Bonzi2011, Bonzi2012}. 
The $L$ X-ray cross sections were measured with monoenergetic excitation beam at 10 keV.

We report cross sections for each spectral line, according to the resolution of the Si(Li) solid state detector used to resolve individual component lines of the spectral emission. The experimental cross sections were grouped 
considering the transitions scheme, the energy of the emission lines and the detector resolution.

In general, the fluorescence cross sections obtained in this work show the same trend with Z and broad agreement with 
the data published by Puri {\it{et al.}} \cite{Puri1993, Puri1995} and Krause \cite{Krause1979, Krause1978}, calculated
 using Scofield's coefficients \cite{Scofield1973, Scofield1974}.

\section{Experimental Condition} 
The measurements were carried out at the X-ray Fluorescence beam line at the 
National Synchrotron Light Laboratory (LNLS), Campinas, Brazil \cite{Perez1998}. The components of the experimental setup were:

\begin{itemize}
\item
 Silicon (111) channel cut double crystal monochromator, which can tune energies between 3 and 30 keV. The energy resolution is 3$\cdot 10^{-4}$ to 4$\cdot 10^{-4}$  between 7 and 10 keV.

	\item  A Si(Li) solid state detector, 5 mm thick and 5 mm in diameter, with a resolution of 170 eV at 5.9 keV and a 0.0127 cm thick beryllium window. The model introduced by Jaklevic and Giauque \cite{Jaklevic1993} was used to obtain the detector efficiency.

	\item  The whole setup is mounted on a motorized lift table, which allows the vertical positioning of the instruments within the linearly polarized part of the beam.

	\item  To limit the beam size, a motorized computer controlled set of vertical and horizontal slits (located upstream and downstream of the monochromator) was used.
\end{itemize}

A set of foil samples (rhodium, palladium, silver, cadmium, indium and tin ) was used to determine the $L$ fluorescence cross sections of these elements.  
The foil samples were provided by Alfa products Inc., with a certified purity of over 99\%. The foils thicknesses are shown in Bonzi {\it{et al.}} (see Table I) \cite{Bonzi2011}.

$K$ emission lines of chlorine, calcium, titanium and iron were measured to determine the geometrical and the detector efficiency factors. 

The $K\alpha$ and $L\alpha$ fluorescent spectra were measured by collecting 2$\cdot10^5$ net counts for each element in order to have the 
same statistical counting error in all measured spectra.

A system dead time, lower than 1\%, was established measuring the fluorescence emission of a Ti sample, adjusting the slit at the exit of the 
monochromator. All samples were measured with the same slit aperture. Unwanted effects, such as piling up, were avoided using this 
configuration and the geometric factors were ensured to be the same for all samples.
 This configuration made it unnecessary to carry out 
corrections for count losses, spectra distortions or modification of the geometrical arrangement.

\section{Spectra analysis}

The energy of the emission lines tabulated by Scofield \cite{Scofield1973, Scofield1974} and the detector resolution were considered to group the $L$ X-ray fluorescence lines. This line arrangement was used to fit the $L$ spectrum, where the $L\beta$ and $L\gamma$ compound lines have been noted with a Roman subscript according to the most intense contribution line, with its corresponding atomic transition:

\begin{itemize}
 \item  $Ll = L_3 - M_1$,

 \item $L\alpha	= L_3 - M_5 + L_3 - M_4$,

 \item  $L\beta_I = L_2 - M_4 + L_1 - M_2 + L_1 - M_3  + L_3 - N_1$,

 \item  $L\beta_{II} = L_3 - N_5 + L_3 - O_4 + L_3  - O_5 + L_3 -$ \\\mbox{}\hspace{12mm}$O_1 + L_1 - M_5 + L_1 - M_4 + L_3 - N_4 $,

 \item  $L\gamma_I = L_2 - N_4$,

 \item  $L\gamma_{II} = L_1 - N_2 + L_1 - N_3 + L_1 - O_2 + L_1 - O_3$.
 
\end{itemize}
     
The background radiation was fitted using a linear second order polynomial.
 
The area of the fluorescence peaks was determined as the average of the areas obtained by the adjustment using Hypermet and Gaussian functions. The escape peaks were fitted using a Gaussian function.

As a consequence of the excitation with a linearly polarized photon beam, the contribution to the background was very low. The linear polarization of the incident beam produces negligible scattered radiation at 90$^0$ with respect to the incident beam direction. The detector position is localized at the same height of the storage ring.

\section{Data Analysis}
     The expression for the $L$ experimental fluorescence cross sections is \cite{Rao1993}

\begin{align}
\sigma^e_{Li}(Eo) = \frac{I_{Li}}{Io.G.\epsilon(E_{Li}).T(Eo,E_{Li})}
\end{align}
where $\sigma^e_{Li}(Eo) = $ experimental $Li$ fluorescence cross sections of the element observed at the energy $Eo$, with $Li = Ll$, $L\alpha$, $L\beta_I$, $L\beta_{II}$, $L\gamma_I$ or $L\gamma_{II}$; $I_{Li} = $ measured intensity of the $Li$ spectral line; $Io.G.\epsilon(E_{Li}) = $ factor comprising the intensity of the excitation beam $Io$; the geometry of the experimental arrangement $G$ and the detector efficiency $\epsilon(E_{Li})$; $Eo = $ energy of the incident beam, in this case 10 keV; $E_{Li} = $ energy of the $Li$ spectral line; the data was obtained from Scofield \cite{Scofield1973}; and $T(Eo,E_{Li}) = $ correction factor for self absorption in an infinitely thick sample, which is

\begin{align}
T(Eo,E_{Li}) = \left(\frac{\mu(Eo)}{\sin(\theta_1)}+\frac{\mu(E_{Li})}{\sin(\theta_2)}\right)^{-1}
\end{align}
where 

$\mu(E) = $ mass absorption coefficient of the sample at energy $E$ from Hubbell and Seltzer \cite{Hubbell1995} and 

$\theta_1$ and $\theta_2 = $ incidence and take off angles, equal to $45^o$ in the current setup.

In these measurements, all the samples were considered as infinitely thick for X-ray fluorescence.

The factor $Io.G.\epsilon(E)$ was calculated using the following expression

\begin{align}
Io.G.\epsilon(E_{Ki}) = \frac{I_{Ki}}{\sigma^{\omega F}_{Ki}(Eo).C_i.T(Eo,E_{Ki})}
\end{align}
where 

$I_{Ki} = $ measured intensity of the $K$ spectral line, 

$Ci = $ the weight concentration of the element of interest in the sample, 

$\sigma^{\omega F}_{Ki}(E) = K $ fluorescence cross sections of the element observed at energy $E$, defined as $\sigma^{\omega F}_{Ki}(Eo) = \sigma_{Ki}(Eo) . \omega_K . F_K$, with $\sigma_{Ki}(E)= K $ shell photoionization cross section for the given element at the excitation energy $E$, from Scofield \cite{Scofield1973},

     $\omega_K = K $ shell fluorescence yield, from Krause \cite{Krause1979,Krause1978} and

     $F_K = $ fractional emission rate for $K\alpha$ or $K\beta$  X-rays, from Khan and Karimi \cite{Khan1980}, defined as

\begin{align}
F_{K\alpha} = \left[1+\frac{I_{K\beta}}{I_{K\alpha}}\right]^{-1} ; 
F_{K\beta} = \left[1+\frac{I_{K\alpha}}{I_{K\beta}}\right]^{-1} 
\end{align}

		 $T(Eo,E_{Ki}) = $ correction factor for self absorption in the sample, 
     $Eo = $ energy of the incident beam and
     $E_{Ki} = $ energy of the $K$ spectral line for a given element, from Scofield \cite{Scofield1973}.

The factor $Io.G.\epsilon(E)$ was previously determined in Ref. \cite{Bonzi2011}, where the same geometry and detector were used. 
Because of this, the $Io.G.\epsilon(E)$ energy dependence is already known and only a scale factor is needed to obtain the correct beam intensity. 

Four targets: Cl (NaCl), Ca (CaHPO$_4$ $\cdot$ 2H$_2$O), Ti (Ti foil) and Fe (Fe foil) emitting fluorescent X-rays
in the range from 2.4 keV to 7.0 keV were used to determine the scale factor in this work. Four $K\alpha$ and four $K\beta$ lines were used to fit the scale factor. Jaklevic and 
Giauque's \cite{Jaklevic1993} model was used to fit the detector efficiency.

\section{Results and Discussion}
\begin{figure}
\begin{center}
\includegraphics[width=0.43\textwidth]{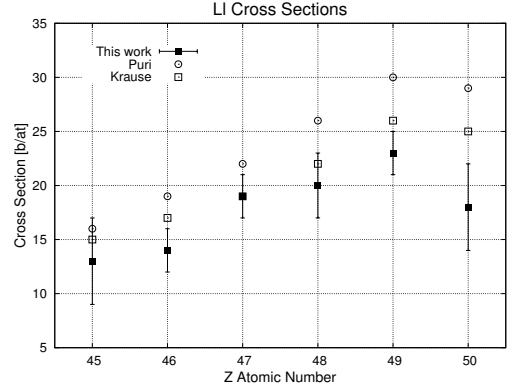}
\end{center}
\caption{Comparison of $Ll$ cross sections.} \label{Lll}
\end{figure}

\begin{figure}
\begin{center}
\includegraphics[width=0.43\textwidth]{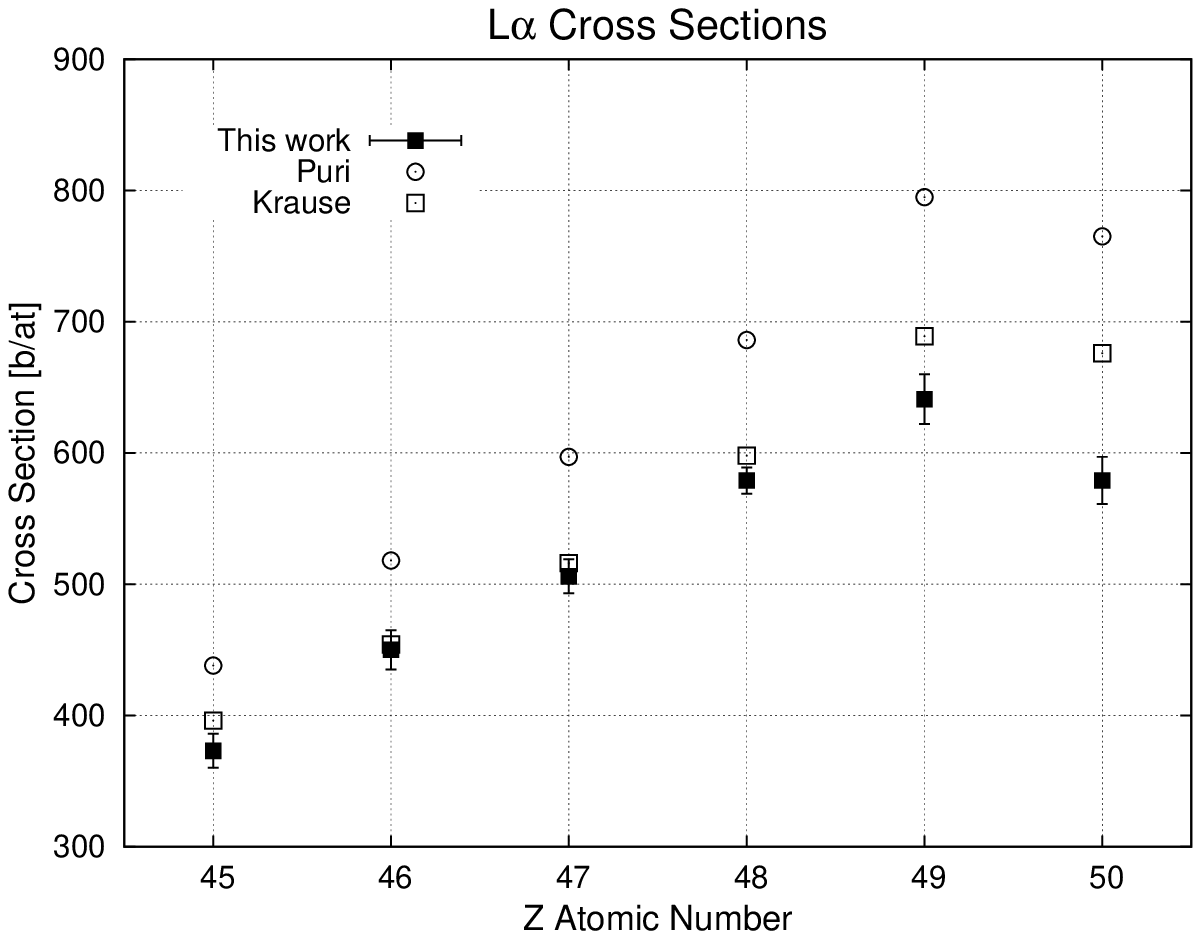}
\end{center}
\caption{Comparison of $L\alpha$ cross sections.} \label{La}
\end{figure}

\begin{figure}
\begin{center}
\includegraphics[width=0.43\textwidth]{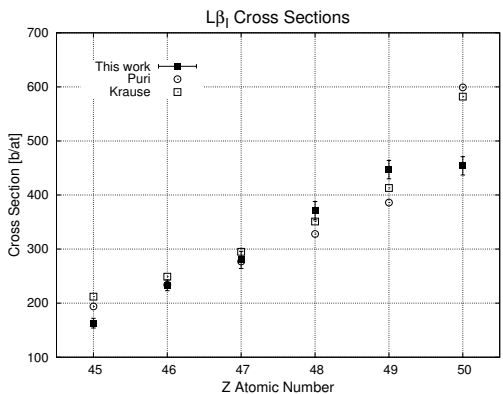}
\end{center}
\caption{Comparison of $L\beta_I$  cross sections.} \label{Lb1}
\end{figure}

\begin{figure}
\begin{center}
\includegraphics[width=0.43\textwidth]{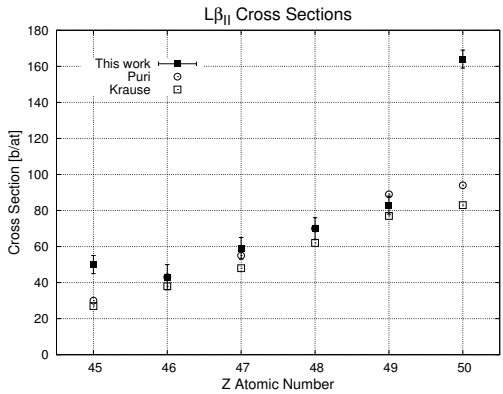}
\end{center}
\caption{Comparison of $L\beta_{II}$ cross sections.} \label{Lb2}
\end{figure}

\begin{figure}
\begin{center}
\includegraphics[width=0.43\textwidth]{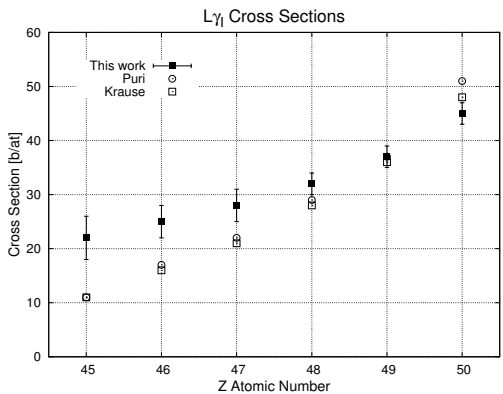}
\end{center}
\caption{Comparison of $L\gamma_I$ cross sections.} \label{Lg1}
\end{figure}

\begin{figure}
\begin{center}
\includegraphics[width=0.43\textwidth]{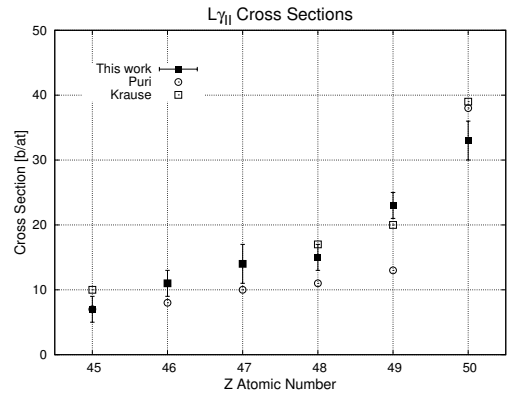}
\end{center}
\caption{Comparison of $L\gamma_{II}$ cross sections.} \label{Lg2}
\end{figure}

$L$ X-ray cross section values obtained in our fluorescence experiment and the theoretical values calculated 
by using coefficients given by Scofield \cite{Scofield1973, Scofield1974}, Puri {\it{et al.}} \cite{Puri1993}
 and Krause \cite{Krause1978} are shown in Table~\ref{Table1} and Figs. \ref{Lll} to \ref{Lg2}.
				
\begin{table*}
\centering
\small
\begin{tabular} [0.75\textwidth]{@{\extracolsep{\fill}}  c  c  c  c  c  c  c  c  }
\hline \hline
Element \ &   \ & $Ll$ \ & $L\alpha$ \ & $L\beta_I$ \ & $L\beta_{II}$ \ & $\gamma_I$  \ & $L\gamma_{II}$ \\
  \hline \hline
         & This work & 13 $\pm$  4 	& 373 $\pm$  13 & 163 $\pm$  9 & 50 $\pm$  5 & 22 $\pm$  4 & 7 $\pm$  2 \\
	 Rh 45 & Puri      & 16        		& 438         	& 194        	& 30        	& 11        	& 7       \\
		  	 & Krause    & 15        		& 396         	& 212        	& 27        	& 11        	& 10      \\
  \hline
         & This work & 14 $\pm$  2 & 450 $\pm$  15 & 233 $\pm$ 10 & 43 $\pm$  7 & 25 $\pm$  3 & 11 $\pm$ 2 \\
	 Pd 46 & Puri      & 19        		& 518         & 234        		& 43        	& 17        	& 8       \\
		  	 & Krause    & 17         	& 454         & 249        		& 38        	& 16        	& 11      \\
	\hline
         & This work & 19 $\pm$  2 & 506 $\pm$ 13 & 280 $\pm$ 16 & 59 $\pm$  6 & 28 $\pm$  3 & 14 $\pm$ 3 \\
	 Ag 47 & Puri      & 22        		& 597        & 277        	& 55        		& 22        	& 10       \\
		  	 & Krause    & 19        		& 516        & 295        	& 48        		& 21        	& 14      \\
 \hline
         & This work & 20 $\pm$  3 & 579 $\pm$ 10 & 371 $\pm$ 17 & 70 $\pm$  6 & 32 $\pm$  2 & 15 $\pm$ 2 \\
	 Cd 48 & Puri      & 26        		& 686        	& 328        		& 70        	& 29         & 11       \\
		  	 & Krause    & 22        		& 598        	& 351        		& 62        	& 28         & 17      \\
\hline
         & This work & 23 $\pm$  2 & 641 $\pm$ 19 & 447 $\pm$ 17 & 83 $\pm$  5 & 37 $\pm$  2 & 23 $\pm$ 2 \\
	 In 49 & Puri      & 30        		& 795        	& 386        		& 89        	& 37         & 13       \\
		  	 & Krause    & 26        		& 689        	& 413        		& 77        	& 36         & 20      \\
\hline
         & This work & 18 $\pm$  4 & 579 $\pm$ 18 & 454 $\pm$ 17 & 164 $\pm$  5 & 45 $\pm$  2 & 33 $\pm$ 3 \\
	 Sn 50 & Puri      & 29        		& 765        	& 599        		& 94        	& 51          & 38       \\
		  	 & Krause    & 25        		& 676        	& 582        		& 83        	& 48          & 39      \\
						
\hline\hline
\end{tabular}
\caption{Experimental and theoretical $L$ X-ray fluorescence  cross sections in Barns/atom at 10 keV. Experimental data (This work), theoretical values calculated using Scofield \cite{Scofield1973} and Puri \cite{Puri1993} and semi-empirical coefficients obtained from Scofield \cite{Scofield1973} and Krause \cite{Krause1979}.}
\label{Table1}
\end{table*}

 Puri {\it{et al.}} predicted theoretical Coster Kronig and fluorescence values using {\it{ab initio}} relativistic
 calculations, while Krause's values of $\omega_K$, $\omega_{Li}$ and $f_{ij}$ were obtained by fitting experimental and
 theoretical compiled data. In Krause's tables, the theoretical data were calculated for singly ionized free atoms while
 the experimental data contain contributions from solid state, chemical and multiple ionization effects. 

The $L\alpha$ cross section values have a better agreement with the theoretical values when the intensity peaks are fitted with an Hipermet function instead of a Gaussian function. 
This happens because the Hipermet function has a tail on the left side that increases the fitted area.

Moreover, the tail of the  Hipermet function used to fit the $L\alpha$ peaks diminishes the area and the cross sections of the $Ll$ peaks, accordingly.

The experimental $Ll$ cross sections show a similar Z trend, compared to the data obtained using Krause and Puri  {\it{et al.}} values. Nevertheless,in general, our results are lower than those.

The $L\alpha$ experimental fluorescence cross section data, Fig. \ref{La}, agree well with Krause's values although
for elements with higher Z, the experimental values are slightly lower than those from Krause. They are even lower than Puri's {\it{et al.}} values, but still showing the same trend with Z.

The $L\beta_I$ experimental fluorescence cross sections show a very good agreement with the theoretical values when  
the Hipermet function is used to fit the area (see Fig. \ref{Lb1}).

The $L\beta_{II}$ measured cross sections show a similar dependence on Z as both theoretical assemblies (Fig. \ref{Lb2}).

The $L\beta_I$ Sn experimental value is lower than the data presented by either Puri {\it{et al.}} 
or Krause and the $L\beta_{II}$ Sn experimental value is much higher than both theoretical values. This behavior might 
be due to the fitting process as both spectra lines are too close in energy; the $L\beta_{II}$ intensity seems to be overestimated while the $L\beta_I$ intensity seems to be underestimated.

A similar behavior is observed for rhodium experimental data although the differences with the theoretical values are 
much smaller than those for tin.

The experimental $L\gamma_I$ fluorescence cross sections show some differences with the Z trend of the theoretical data: in the lower Z range, the experimental values are higher than the theoretical ones while for higher Z,
 this difference becomes smaller. Sn values, Z = 50, show a different behavior, being lower than both calculated values (see Fig. \ref{Lg1}).

 The $L\gamma_{II}$ experimental values show the general Z trend of the values presented by Krause and Puri {\it{et al.}}.
 The experimental values are sometimes higher or lower than the theoretical ones but the range of values 
is similar to them (see Fig. \ref{Lg2}).

To determine the uncertainties of the experimental cross sections, the propagation of errors was carried out in Eq. (1).
 The uncertainty values are in general around 6-10\%, and less than 40\% in case of the $Ll$ line.

The uncertainty associated to the $Io.G.\epsilon(E)$ factor was estimated as the mean quadratic deviation of 
the experimental values ($\leq2$\%). For the factor $T(Eo,E_{Li})$, a propagation of errors was carried out
assuming a 3\% error in the values of the mass absorption coefficients, and a 2\% error in the sine of the angles 
due to the sample positioning errors.
Krause's $\omega_K$ values for elements with 45 $\leq$ Z $\leq$ 50 have an estimated error of 1\%.

The uncertainties of the peak areas were established as half the difference between the areas obtained using Gaussian and Hypermet functions to fit. These uncertainties were the main contribution to the experimental errors of the cross section.

\section{Conclusions}

In this investigation, the $L$ X-ray fluorescence cross sections of a group of elements with 45 $\leq$ Z $\leq$ 50
were measured using a synchrotron radiation source for monoenergetic beams at 10  keV. The polarization
properties of the monoenergetic excitation beam and the high resolution of the detector system allowed to
reduce the scattered radiation thus obtaining a better signal to noise ratio and a better accuracy 
for the experimental cross sections.

The cross sections of $Ll$, $L\alpha$, $L\beta_I$, $L\beta_{II}$, $L\gamma_I$ and $L\gamma_{II}$ lines were measured 
considering a more detailed group than the usual sets.
In Table~\ref{Table1}, the comparison between the experimental fluorescence cross section values with the theoretical values
 calculated using coefficients from Scofield \cite{Scofield1973,Scofield1974}, Puri {\it{et al.}} \cite{Puri1993} and 
Krause \cite{Krause1979} are shown.

Our experimental values are in general in good agreement with the calculated data 
using Scofield's \cite{Scofield1973, Scofield1974} and Krause's \cite{Krause1979} coefficients.

The $L$ cross sections present uncertainties around 6-10\% and the less intensive $Ll$ peaks
show uncertainties that in some cases come close to 40\%, being the fitting uncertainty the most 
important error source. 

The use of the Hypermet function is very convenient to fit the $L\alpha$ and $L\beta$ peaks (see Table~\ref{Table1}).

The solid state detector used in our experiments does not have enough energy resolution to resolve each spectral line. A higher resolution detection system would be desirable in order to analyze each spectral line separately.

The Coster Kronig coefficients present large fluctuations in this atomic range and that is the cause of the
 observed discrepancies. 

\begin{acknowledgements}
This work was carried out under grants provided by SeCyT U.N.C. (Argentina). 
Research partially supported by LNLS - National Synchrotron Light Laboratory, Brazil.
\end{acknowledgements}

\end{document}